\documentclass[runningheads]{svmult}

\usepackage{makeidx}   
\usepackage{graphicx}  
\usepackage{subeqnar}  
\usepackage{multicol}  
\usepackage{physprbb}  
\makeindex             

\newcommand{\be}{\begin{equation}}
\newcommand{\ee}{\end{equation}}
\newcommand{\bea}{\begin{eqnarray}}
\newcommand{\eea}{\end{eqnarray}}

\newcommand{\sla}{\raise.15ex\hbox{$/$}\kern-.57em}
\begin{document}
\title*{Lattice chiral fermions from continuum defects}
\toctitle{Lattice chiral fermions from continuum defects}
%
%
\titlerunning{Lattice chiral fermions from continuum defects}
%
\author{Herbert Neuberger\inst{1,}\inst{2}}

\authorrunning{Herbert Neuberger}
%
%
\institute{School of Natural Sciences,\\
           Institute for Advanced Study, Princeton, NJ 08540
\and Rutgers University, \footnote{Permanent Address.}\\
Department of Physics and Astronomy\\
Piscataway, NJ 08855, USA \\
}

\maketitle              

\begin{abstract}
We consider whether defects of co-dimension two
could produce new lattice chiral fermions.
\end{abstract}

\section{Introduction}

A vector like gauge theory with an  
infinite number of flavors can have a mass matrix that
has an index equal to unity, but otherwise is of the order of the
ultra-violet cutoff~\cite{plb1}. This produces a 
strictly massless charged Weyl
particle accompanied by an infinite number of heavy Dirac particles. 
This setup has been used on the lattice
with a domain-wall~\cite{kapplb} realization of the mass matrix 
where it leads to the overlap~\cite{review}.
Here we ask whether a string realization of infinite 
flavor space might achieve a substantially different form of
lattice chirality. 
The question is left unanswered, but there is some indication
that employing a variant of Baxter's corner transfer matrix (CTM)
~\cite{baxter} a more
general construction of lattice chiral fermions 
might be possible. It is hoped that the present discussion
will be a motivator to develop the idea further. 

\section{Infinite flavor space}

Many reviews have been written about the new way of putting chiral
fermions on the lattice - for one example see~\cite{review}. 
Therefore, after a very brief overview, 
this section will mention only facts 
essential for what follows later. 

The overlap construction was based on two independent suggestions
about regularizing chiral gauge theories, one by D. B. Kaplan~\cite{kapplb}
and the other by S. Frolov and A. Slavnov~\cite{froslav}. The first of
these was, in turn, motivated by an earlier paper by 
C. Callan and J. Harvey~\cite{ch}.
This earlier paper is also the basis of the present article.
The overlap path of development was completed in the summer of 1997
with the publication of~\cite{ovlap}. 

One very often hears in this context about a paper 
by Ginsparg and Wilson~\cite{gw}, published in 1982, 
which fell into oblivion 
until after the publication of the first paper in~\cite{ovlap}.
In the second paper in ~\cite{ovlap} the overlap
line of development was merged with the Ginsparg-Wilson proposal.
This confluence played an important psychological role in
convincing the majority of workers in the field
that indeed the problem of lattice chirality was solved by the overlap. 
At the moment there exists only one construction 
of lattice chirality that truly 
works. To understand this construction 
as well as our attempt to go beyond it there is
no need to be familiar with ~\cite{gw} (although
this paper is strongly recommended on general grounds) 
nor with the many papers re-deriving the overlap
with the Ginsparg-Wilson relation as a a starting point. At the practical
level too, it is fair to say, little concrete gain has been produced by 
the many Ginsparg-Wilson based papers appearing from 1998 onwards, 
except a sobering recognition that the
problem of lattice chirality could have been solved in 1982, or soon
after, had we focused on this prescient paper. 

In this article we shall not diverge very much from what has proven to be
successful. We simply go to another example in the paper of Callan and Harvey,
and using yet another observation of D. B. Kaplan~\cite{kap} try to follow the 
procedure undertaken when the overlap was developed to see where this
ends up leading us. To do this it is necessary
to only review the basic logical steps in the overlap construction. 

The general framework that unites the ideas of D. B. Kaplan~\cite{kapplb} and 
of S. Frolov and A. Slavnov~\cite{froslav} is to think about a formally vector-like
gauge theory which contains an infinite number of flavors~\cite{plb1}. The main
new object in the Lagrangian is a mass matrix that has an analytic index
equal to unity and whose non-zero spectrum is separated from zero
by a gap of order the cutoff energy:
\be
\label{mother}
{\cal L}_\psi=
\bar\psi \gamma_\mu D_\mu \psi +\bar\psi_R M \psi_L
+\bar\psi_L M^\dagger \psi_R
\ee

The conditions on $M$ are:
\begin{itemize}
\item $M$ has exactly one zero mode.
\item $M^\dagger$ has no zero mode. (The adjoint of $M$ is defined so that
the Euclidean fermion propagators have the right formal adjointness properties.)
\item The spectrum of $MM^\dagger$ is separated from zero by a finite gap.
\end{itemize}
If these condition hold for $M$, they will also hold for $M+\delta M$,
if $\delta M$ is small enough. This property protects the mechanism from
being destroyed by radiative corrections. 

These conditions ensure that there will be one Weyl fermion and
an infinite number of heavy Dirac fermions. 
To make this into a well defined construction - one that can
be implemented on a finite computer - the following steps
were carried out in the context of the 
domain-wall/overlap construction:
\begin{itemize}
\item{Four dimensional space-time is replaced by a finite
lattice. Invariably one chooses a torus shape. All ordinary
infinities are eliminated by this.}
\item{In order to deal with the single remaining infinity (in flavor space) one makes 
as simple a choice as one can for $M$ and the space it acts on.}
\item{One next formally integrates out all fermions. This step
is formal because the number of fermions is infinite. Nevertheless,
for special choices of $M$, one can interpret the result
as something finite and well defined times an infinite factor
that looks, intuitively, harmless. The harmless infinite factor
is discarded and the remaining structure is a well defined
candidate for the chiral determinant line bundle over gauge
configuration space.}
\item{The candidate is subjected to several tests: it has to
reproduce instanton physics and the associated fermion number 
conservation violating processes; it has to reproduce anomalies 
of global and local type, etc. (It goes without saying that ordinary
Feynman diagrams must be reproduced.)}
\item{In the vector-like case one
can next try to derive as simple a form as possible for the effective
action and see directly how continuum chiral Ward identities are reproduced
by this action.}
\end{itemize}
We shall go some way on this path in the case of a string defect
\footnote{Although we do not get very far, we already find 
ourselves in disagreement with a recent paper by Nagao.}.

\section{Two dimensional flavor space}

\subsection{Axial symmetric case}

Let us start by reviewing Callan and Harvey's paper,
~\cite{ch} - and add some further details, mainly for pedagogical reasons.

First we need to set our conventions for the Gamma matrices in 
6 Euclidean dimensions. The various directions will be labeled as: 
$a=1,2,3,4,5,6$, $\mu=1,2,3,4$, $\alpha,\beta=5,6$.

\be
\Gamma_\mu=\sigma_1\otimes \gamma_\mu~~~~~
\Gamma_5 =\sigma_1\otimes \gamma_5~~~~~
\Gamma_6=\sigma_2 \otimes 1
\ee
As a result,
\be \Gamma_a = \pmatrix{0&\rho_a\cr \rho_a^\dagger &0}
\ee
\be
\Gamma_7 = -i
\Gamma_1\Gamma_2\Gamma_3
\Gamma_4\Gamma_5\Gamma_6=-i(1\otimes\gamma_5)(i\sigma_3\otimes\gamma_5)=\sigma_3\otimes 1
\ee
or
\be
\Gamma_7 = \pmatrix {1&0\cr 0&-1}
\ee
The Weyl matrices are:
\be
\rho_\mu=\gamma_\mu,~~~~~\rho_5=\gamma_5,~~~~~\rho_6=i
\ee
Thus, for 6D chiral fermions we can use four component spinors with action
\be
{\cal L}_\psi=\bar\psi (\rho_a^\dagger D_a ) \psi
\ee
{}From the 4D point of view one can think about the $\psi$'s as Dirac fermions
consisting of two 4D Weyl fermions of opposite handedness $\psi_{L,R}$:
\be 
\gamma_5\psi_L=\psi_L,~~\gamma_5\psi_R=-\psi_R,~~
\bar\psi_R \gamma_5=\bar\psi_R,~~\bar\psi_L\gamma_5 =-\bar\psi_L
\ee
Each one of the above four fermion fields has two unrestricted components. 
\be
{\cal L}_\psi = \bar\psi \gamma_\mu D_\mu \psi +\bar\psi_R (iD_6+D_5)\psi_L
+\bar\psi_L (iD_6 - D_5 )\psi_R
\ee
With $m=iD_6 +D_5$ and $m^\dagger=iD_6 -D_5$ 
this can then be written in the generic form:
\be
{\cal L}_\psi = \bar\psi\sla D_4 \psi+ \bar\psi_R {\it m} \psi_L +
\bar\psi_L {\it m}^\dagger \psi_R 
\ee
${\it m}$, viewed as an operator on the 2D space of $x_5$ and $x_6$,
has a hermitian conjugate, ${\it m}^\dagger$. The $x_\alpha,~\alpha=5,6$
are viewed as continuous indices in flavor space. 

In the Callan-Harvey set-up we have, in six dimensions, one Dirac fermion, 
comprising of a four
component $\psi$ as above, as well as another field, 
a four component $\chi$, 
of opposite handedness.
\bea
&{\cal L}_{\psi,\chi} = \bar\psi \sla D_4 \psi + \bar\chi \sla D_4 \chi +\\
\nonumber
&\pmatrix { \bar\psi_R & \bar \chi_R }
\pmatrix { {\it m} & \Phi\cr \Phi^* & - {\it m}^\dagger} 
\pmatrix{\psi_L \cr \chi_L } +\\
\nonumber
&\pmatrix { \bar\psi_L & \bar \chi_L }
{\pmatrix { {\it m} & \Phi\cr \Phi^* & - {\it m}^\dagger}}^\dagger 
\pmatrix{\psi_R \cr \chi_R }\\\nonumber 
\eea
We see that the structure has the generic form of~(\ref{mother})  
with a mass matrix $M$ given by
\be
M=\pmatrix{{\it m}& \Phi\cr \Phi^* & -{\it m}^\dagger }
\ee
and, in terms of the flavor doublet $\Psi\equiv\pmatrix{\psi\cr \chi}$, 
a fermionic Lagrangian of the form
\be
{\cal L}_\Psi = \bar\Psi (1\otimes \sla D_4 )\Psi + 
\Psi_R (M\otimes 1) \Psi_L + \Psi_L (M^\dagger\otimes 1) \Psi_R
\ee
Above, $\psi$ and $\chi$ are viewed as two four dimensional Euclidean Dirac
fermions and are distinguished by the discrete portion of a flavor index. This 
discrete portion can take one of two possible values. 
The unit matrices in the direct products are all two by two
while $\sla D_4$ is four by four and $M$ is two by two. In this
dimension-counting we have ignored all color indices and the continuum labels. 

$M$ is arranged to have unit analytical index by picking
\be
\Phi = e^{i\phi} f(r)
\ee
where $x_5 = r\cos\phi$ and $x_6=r\sin\phi$. 
The natural inner product has an extra factor of 
$r$ in the measure. As a result, 
$\partial_r^\dagger = -\partial_r -\frac{1}{r}$. 
Taking the $A_5$ and $A_6$ gauge 
field components to vanish we have
\bea
{\it m} = \partial_5 + i \partial_6 = e^{i\phi}
 (\partial_r +\frac{i}{r}\partial_\phi )\nonumber\\
{\it m}^\dagger = -\partial_5 + i \partial_6 = e^{-i\phi}
 (-\partial_r +\frac{i}{r}\partial_\phi )
\eea
The matrix $M$ now becomes
\be
M=\pmatrix{ e^{i\phi} & 0 \cr 0 & e^{-i\phi}} 
\pmatrix {\partial_r +\frac{i}{r} \partial_\phi & f(r)\cr
f(r) & \partial_r -\frac{i}{r} \partial_\phi}
\ee
leading to
\be
M^\dagger = \pmatrix { -\partial_r -\frac{1}{r} +\frac{i}{r} \partial_\phi & f(r)
\cr f(r) &  -\partial_r -\frac{1}{r} - \frac{i}{r} \partial_\phi }
\pmatrix{ e^{-i\phi} & 0 \cr 0 & e^{i\phi}} 
\ee

\subsubsection{$M$ has a single zero mode. }

We now look for zero eigenstates of $M$. 
The zero mode equation is
\be
\pmatrix {\partial_r -\frac{k}{r}  & f(r)\cr
f(r) & \partial_r +\frac{k}{r} } \pmatrix{\tilde\eta_{1k}\cr\tilde\eta_{2k}}=0
\ee
Zero modes of $M$ will be two component functions of $x_5$ and $x_6$, denoted by  
$\eta$. In the $r,\phi$ polar coordinates, in order 
to get a zero mode, both
components of $\eta$ need to carry the same angular momentum $k\in Z$.
Thus, the angular dependence of $\eta$ is of the form $e^{ik\phi} \tilde\eta$
where $\tilde \eta$ is a function of $r$ only. 
A zero mode of $M$ is also a zero mode of $M^\dagger M$.
Acting on an $\tilde\eta$ field carrying angular momentum $k$, 
$M^\dagger M$ has the form
\be
M^\dagger M = \pmatrix { -\partial_r -\frac{1}{r} -\frac{k}{r}& f
\cr f &  -\partial_r -\frac{1}{r} +\frac{k}{r}}
\pmatrix {\partial_r -\frac{k}{r} & f\cr
f & \partial_r +\frac{k}{r}}
\ee

A bit of algebra gives
\be
\label{m-dag-m}
M^\dagger M = 
\pmatrix{-\partial_r^2 -\frac{1}{r}\partial_r +\frac{k^2}{r^2} + f^2 &
-f^\prime  -\frac{f}{r} \cr
-f^\prime  -\frac{f}{r} &
-\partial_r^2 -\frac{1}{r}\partial_r +\frac{k^2}{r^2} + f^2}
\ee
We can go to the sum and difference 
fields $\frac{1}{\sqrt{2}}(\psi\pm\chi)$ 
which diagonalizes $M^\dagger M$ to:
\bea
&\frac{1}{2}\pmatrix{1&1\cr 1&-1} M^\dagger M 
\pmatrix{1&1\cr 1&-1}=\\\nonumber
&\pmatrix{-\partial_r^2 -\frac{1}{r}\partial + f^2 - f^\prime -\frac{f}{r}
+\frac{k^2}{r^2} & 0 \cr
0 & -\partial_r^2 -\frac{1}{r}\partial + f^2 + f^\prime + \frac{f}{r}
+\frac{k^2}{r^2}}
\eea
This can also be written as:
\bea
&\frac{1}{2}\pmatrix{1&1\cr 1&-1} M^\dagger M \pmatrix{1&1\cr 1&-1}=
\nonumber\\
&\pmatrix { (\partial_r + f)^\dagger (\partial_r +f ) & 0\cr
0&(\partial_r - f)^\dagger (\partial_r - f  )}+\frac{k^2}{r^2}
\pmatrix{1&0\cr 0&1}
\eea

The two terms above are both positive semi-definite operators and, 
for a zero
mode, they must both vanish on $\tilde\eta$. This requires $k=0$. 
The zero mode will have to solve the following equation:
\be
\pmatrix{\partial_r +f(r) & 0\cr 0 & \partial_r-f(r)}
\pmatrix {\tilde\eta_1+\tilde\eta_2\cr\tilde\eta_1-\tilde\eta_2}=0
\ee
Only the operator in the left-upper entry has a zero mode, given
by $e^{-\int^r_0 f(\rho) d\rho}$. We conclude that $M$ has a single zero modes
in all cases of interest. 

\subsubsection{$M^\dagger$ has no zero mode. }

We would like now to prove that $M^\dagger$ has no zero modes.
We first go to $MM^\dagger$ hoping to show it is strictly positive.
\bea
&MM^\dagger\pmatrix{\psi\cr \chi}=\\\nonumber
&\pmatrix{-\partial_r^2 -\frac{1}{r}\partial_r -\frac{1}{r^2} 
\partial_\phi^2 + f^2 & (f^\prime -\frac{f}{r} ) e^{2i\phi}\cr
(f^\prime -\frac{f}{r} ) e^{-2i\phi} & 
-\partial_r^2 -\frac{1}{r}\partial_r -\frac{1}{r^2} 
\partial_\phi^2 + f^2 }
\pmatrix{\psi\cr \chi}
\eea
Expanding in angular momentum modes,
\be
\psi=\sum_n e^{in\phi} \psi_n (\phi),~~~~
\chi=\sum_n e^{in\phi} \chi_n (\phi)
\ee
we obtain for the zero mode equation
\bea
&[-\partial_r^2 -\frac{1}{r}\partial_r +
\frac{(n+1)^2}{r^2} + f^2]\psi_{n+1}
+ (f^\prime -\frac{f}{r} )\chi_{n-1}=0\\\nonumber
&[-\partial_r^2 -\frac{1}{r}\partial_r +\frac{(n-1)^2}{r^2} 
 + f^2]\chi_{n-1}+
(f^\prime -\frac{f}{r} )\psi_{n+1}=0
\eea
Note that the ordered pair $(\chi_{-n-1},\psi_{-n+1})$
satisfies the same equations as the ordered 
pair $(\psi_{n+1},\chi_{n-1})$, so one could restrict
the analysis to only $n\ge 0$. 

In matrix notation, we are dealing with the operator
\be
O=\pmatrix{ -\partial_r^2 -\frac{1}{r}\partial_r +
\frac{(n+1)^2}{r^2} + f^2 & f^\prime -\frac{f}{r}\cr
f^\prime -\frac{f}{r} & -\partial_r^2 -\frac{1}{r}\partial_r +
\frac{(n-1)^2}{r^2} + f^2 }
\ee
Consider the identity
\be
(\partial_r + f ) (\partial_r +f)^\dagger =
-\partial_r^2 -\frac{1}{r}\partial_r +\frac{1}{r^2}+f^\prime -
\frac{f}{r}+f^2
\ee
Let us take $n=0$ first. Then we find, with 
\be
K=\frac{1}{\sqrt{2}} \pmatrix{1&1\cr 1& -1},
\ee
\be
KOK= 
\pmatrix { (\partial_r +f)(\partial_r+f)^\dagger & 0\cr
0 & (\partial_r -f ) (\partial_r -f)^\dagger }
\ee
It is easy to see that $KOK$ has no zero modes.
For $|n|\ge 2$ the operator $O$ can be written as the
operator $O$ for $n=0$ plus a non-negative operator.
Hence, we have established that $MM^\dagger$ has no
zero modes with $n=0$ and with $|n|\ge 2$. The case
$|n|=1$ needs special analysis. Due to the symmetry 
between $n=+1$ and $n=-1$
we know that $M^\dagger$ either has no zero modes, 
or that it has two zero
modes. In either case, $M$ has a nonzero index.

With some mild additional assumptions on $f$ 
we can deal with the $n=1$ case too.
These assumptions are not necessary; in other words, 
one can eliminate the case $n=1$ also with different assumptions.

We assume that $f$ satisfies:
\be
\label{cond-on-f}
f^\prime \ge 0,~~~~~\left (\frac{f}{r}\right )^\prime = \frac{1}{r}
\left (f^\prime -\frac{f}{r}\right )\le 0,~~~~~f(\infty)=v>0
\ee
For example, an $f$ given by
\be
f(r)=\frac{vr}{r+a},~~~~~~a>0
\ee
satisfies these assumptions. Actually, $f(0)=0$ should hold always,
since the phase of the field $\Phi$ winds around the origin and
$|\Phi|=f$. Another example would be
\be
f=v \tanh\frac{r}{a},~~~f^\prime-\frac{f}{r} =
\frac{\frac{1}{a}-\frac{1}{2r}
\sinh\frac{2r}{a}}{\cosh^2\frac{r}{a}}
\ee 
The expression is indeed negative 
because $\sinh x=x+\frac{x^3}{3\!}+
\dots$ . 

The assumptions on $f$ (\ref{cond-on-f}) 
are satisfied for all non-zero values of the dimensionless
parameter $\lambda$, defined by:
\be
\lambda=va
\ee

Consider now the identities:
\bea
&(\partial_r+f)(\partial_r+f)^\dagger=-\partial_r^2-\frac{1}{r}\partial_r +
f^2 +f^\prime -\frac{f}{r} +\frac{1}{r^2}\\\nonumber
&(\partial_r+f)^\dagger (\partial_r+f) =-\partial_r^2-\frac{1}{r}\partial_r +
f^2 -f^\prime -\frac{f}{r} 
\eea
Using them we write
\bea\nonumber
&\pmatrix{-\partial_r^2 - \frac{1}{r}\partial_r +\frac{2}{r^2}+f^2 &
f^\prime -\frac{f}{r}\cr
f^\prime -\frac{f}{r} & -\partial_r^2 -\frac{1}{r}\partial_r + f^2 }=
\\\nonumber
&\pmatrix{\frac{1}{r^2} +(\partial_r +f ) (\partial_r +f)^\dagger - f^\prime 
+\frac{f}{r} & f^\prime -\frac{f}{r}\cr
f^\prime -\frac{f}{r} & (\partial_r+f)^\dagger (\partial_r+f)+f^\prime
+\frac{f}{r}}
\\\nonumber
&=\frac{1}{r^2} \pmatrix{1&0\cr 0&1} +(f^\prime -\frac{f}{r})\pmatrix{
-1&1\cr 1 & -1} +\frac{2f^\prime}{r} \pmatrix{0&0\cr 0&1}+
\\
&\pmatrix{(\partial_r+f)(\partial_r+f)^\dagger&0\cr
0&(\partial_r+f)^\dagger(\partial_r+f)}
\eea
Since the matrix
\be
\pmatrix{-1&1\cr 1 & -1}
\ee
is negative semi-definite, every term in the above additive decomposition
is positive semi-definite. For a zero mode they all have to have zero
expectation value and this is impossible.

A similar decomposition should handle the case
$(\frac{f}{r})^\prime \ge 0$ which would correspond to an $f$
which blows up at infinity. 

\subsubsection{Spectral gap. }

Let
\be
f(r)=vg_\lambda (x),~~{\rm with}~~\lim_{\lambda\to 0}g_\lambda (x) = 1
\ee
where $x=vr$ and $\lambda$ is the positive dimensionless parameter introduced
earlier.

To establish the existence of a gap we need to show that the spectrum of
$M^\dagger M$ is separated by a gap from its single zero eigenstate. 
It suffices to show this for wave-functions with no $\phi$ dependence,
see (\ref{m-dag-m}). The non-zero portion of the spectrum of $M^\dagger M$ also
is the joint spectrum of $(\partial_r + f)(\partial_r + f)^\dagger$
and $(\partial_r - f)^\dagger(\partial_r - f)$. We use 
$(\partial_r + f)(\partial_r + f)^\dagger$ rather than 
$(\partial_r + f)^\dagger (\partial_r + f)$ to eliminate the zero mode.

\bea
& \partial_r \partial_r^\dagger = \partial_r^\dagger \partial_r 
+\frac{1}{r^2}\\\nonumber
&(\partial_r + f)(\partial_r + f)^\dagger =\partial_r^\dagger \partial_r 
+ f^2 +f^\prime -\frac{f}{r} +\frac{1}{r^2}\\\nonumber
&(\partial_r - f)^\dagger(\partial_r - f)=
\partial_r^\dagger \partial_r 
+ f^2 - f^\prime +\frac{f}{r} 
\eea
So, all we need is to find lower bounds for 
\bea
h_1=g_\lambda^2+g_\lambda^\prime-\frac{g_\lambda}{x} +\frac{1}{x^2}\\\nonumber
h_2=g_\lambda^2-g_\lambda^\prime +\frac{g_\lambda}{x}
\eea

For $\lambda\to 0$ $g_\lambda(x) \to 1$ and it is easy to check that
$h_{1,2}$ are bounded by positive numbers:
\bea
&h_1 \to \frac{1}{x^2}-\frac{1}{x} +1 \ge \frac{3}{4}\\\nonumber
&h_2=1+\frac{1}{x} \ge 1
\eea

Making $\lambda$ positive nonzero and small cannot change these bounds by much,
for any $x$ away from zero. It is easy to see that at $x=0$, where smoothness 
in $\lambda$ breaks down, these positive lower bounds still hold. Thus,
keeping $\lambda$ small enough we are sure we achieved all our objectives.

\subsubsection{Effective low energy theory. }

We could restrict the fields $\psi_L$, $\chi_L$, $\bar\psi_L$, 
$\bar\chi_L$ to have no dependence on $\phi$, the $\phi$
dependence of $\psi_R$, $\bar\chi_R$ to be given by $e^{i\phi}$
and the $\phi$ dependence of $\chi_R$, $\bar\psi_R$ to be
given by $e^{-i\phi}$. This requires us to constrain
the fields $\bar\psi_R$, $\psi_R$, $\bar\chi_R$, $\chi_R$ 
to vanish at the origin, where the angle $\phi$ is not defined.
After extracting the angular dependence we
can drop the $\phi$ integral in the action 
and again rotate the new $\psi$,$\chi$ fields to $\psi\pm\chi$.
The difference combination can also be dropped.
All the fields we have dropped are ``heavy'' and dropping them
is consistent with effective field theory logic. The scale of their
masses is given by the asymptotic value of $f$ at infinity.
The remaining heavy and light fields (from the four dimensional
point of view) correspond to a single, four component, complex, 
fermion field living in a five dimensional world with a boundary
at $r=0$, where $r$ is the fifth dimension. 
The single zero mode we have is left-handed from the four
dimensional point of view. 

We end up with a set-up similar to the domain wall
case, only that the extra dimension is now $[0,\infty)$ and there is 
an $r$-factor in the internal measure factor. 
Because of this similarity
it is unlikely that a truly new may 
of regularizing chiral fermions could be obtained pursuing this any 
further.

\subsection{A less symmetric arrangement}
David Kaplan ~\cite{kap} had the idea to make a Cartesian arrangement in which the 
phase of $\Phi$ changes in jumps between consecutive constant values
in each quarter of plane. The length of $\Phi$ is frozen to $v$.
One way to do this is to take the scalar field as:
\be
\Phi(x_5, x_6 ) = v[\epsilon(x_5)+i\epsilon(x_6)]
\ee
The sign functions can be smoothed out at the origin. This would
also smooth out the discontinuities in the phase - $\Phi$ would become 
a smooth function. Now
\be
\label{mdisc}
M=\pmatrix{ i\partial_6 + \partial_5  & v[\epsilon (x_5)+i\epsilon (x_6)]\cr
v[\epsilon (x_5)-i\epsilon (x_6)] & -i\partial_6 + \partial_5}
\ee
The zero mode equations become
\bea
[i\partial_6 + \partial_5]\eta_1 + v[\epsilon (x_5)+i\epsilon (x_6)]\eta_2=0\\
\nonumber
[-i\partial_6 + \partial_5]\eta_1 + v[\epsilon (x_5)-i\epsilon (x_6)]\eta_2=0\\
\nonumber
\eea
The zero mode is unique, and given by
\be
\eta_1=\eta_2=Ae^{-v[|x_5|+|x_6|]}
\ee
This mode is localized at the origin, which is the point
at which the two lines on which $f_5$ and $f_6$ vanish
intersect. 
In the original Callan-Harvey setting the two lines 
which intersect are lines
on which the real part and, respectively, the 
imaginary part of the complex
frozen Higgs field, vanish. 

If we smooth out the sign functions, we shall have a differentiable
zero mode. This may not be necessary when we go to a lattice. 
The arrangement allows us  to go to a regular square lattice in
the $x_5 - x_6$ plane. The lattice will have to be
infinite. If we truncate it, on its boundary
we expect to find a partner to the chiral fermion at the origin,
of opposite chirality. 

Our considerations so far put us
in disagreement with Nagao's paper ~\cite{nagao}: We do not
find extra zero modes in the continuum beyond the one four
dimensional Weyl fermion
at the string defect
and we see no need for an axially symmetric lattice 
structure which is necessarily inhomogeneous. Only
the winding of the $\phi$ phase is necessary for the
Callan-Harvey mechanism to work, and that we have shown 
can be achieved by a Cartesian arrangement.

\subsection{Smooth Cartesian arrangement}

We now eliminate all jumps in the Cartesian case:

\be 
v\epsilon(x_\alpha) \to f_\alpha (x_\alpha )
\ee

The real functions  $f_\alpha$ depend on a single real argument
on the infinite real line. They are monotonically increasing and asymptotically
constant, going from $-v_\alpha^{-}$ at $-\infty$ to
$v_\alpha^{+}$ at $\infty$, with $v_\alpha^\pm > 0$ and large, of the
order of a 4D ultraviolet cutoff. Also, we choose $f_\alpha (0)=0$.

The scalar field $\Phi$ is:
\be
\Phi (x_5, x_6 ) = f_5 (x_5 ) +i f(x_6 )
\ee

Define two operators on functions of a $x_\alpha$:
\be
d_\alpha = \partial_\alpha + f( x_\alpha ),~~~~
d_\alpha^\dagger = -\partial_\alpha + f( x_\alpha ),~~~~[d_5 , d_6 ]=0
\ee
The $d_\alpha$'s have zero modes but the $d_\alpha^\dagger$ do not.

Change again the field basis by 
$\psi,\chi\to \frac{1}{\sqrt{2}} (\psi \pm \chi )$ and the same for
$\bar\psi,\bar\chi$. The Lagrangian changes by changing the matrix
$M$:
\be
M\to \frac{1}{2} \pmatrix{1 & 1\cr 1& -1} M  \pmatrix{1 & 1\cr 1& -1} =
\pmatrix {d_5 & -id_6^\dagger \cr id_6 & -d_5^\dagger }
\ee
We shall continue to denote the (new) mass matrix by $M$. 

\subsubsection{Properties of $M$ in the Cartesian case. }

We now need to prove that indeed $M$ has a single zero mode and that
$M^\dagger$ has no zero mode. This is very easy:
\be
M^\dagger M = \pmatrix {d_5^\dagger d_5 + d_6^\dagger d_6 & 0 \cr
0& d_5 d_5^\dagger + d_6 d_6^\dagger }
\ee
\be
MM^\dagger =\pmatrix {d_5 d_5^\dagger + d_6^\dagger d_6 & 0 \cr
0& d_5^\dagger d_5 + d_6 d_6^\dagger }
\ee
This not only proves our statement, but also shows that the 
zero eigenvalue of $M^\dagger M$ is separated by a large gap from the
the rest of the spectrum. 
Hence, the arrangement is stable under deformations smaller
than the cutoff scale.

Note that the above construction is a general prescription of combining
two commuting operators with index into a new operator with the same
index acting on a doubled product space. This procedure
can be iterated, doubling the number of extra dimensions at each step.

\subsubsection{Spectral gap in the Cartesian case. }

Here we deal with the two functions $f_{1,2}$ which 
obey $f^2_\alpha +f^\prime_\alpha \to v^2$ as $\lambda$, approaches
zero and the same logic as in the previous gap analysis, for the axially 
symmetric case, again applies. 

\section{Compactifications}

If the 2D $x_5 - x_6$ plane is compactified the chiral fermion will 
get a partner of opposite chirality and the index will be lost.
Actually, one can generate more  
Dirac fermions than the minimum needed, as we shall see.

Consider first the axially symmetric case: 
To obtain a compactification we need to replace $f$ by a
periodic function.  $f(r)$ would go from zero at $r=0$ to
zero at $r=R$. It would grow to a positive large value, 
stay at it for a long while, and then drop back to zero at $R$. 
It is natural to compactify now to a sphere. Each half sphere
contains one of the poles of the original sphere and a circle on the
boundary where $f$ is in the middle of its range. At each
pole we have a chiral fermion, the chiralities at the two poles
being opposite. In total, we have a Dirac fermion. 

Turn now to the Cartesian case:
$f_\alpha$ is made periodic as follows: it would go from 
$-v_\alpha^-$ at zero to $ v_\alpha^+$ and continue back after a long while
to $-v_\alpha^-$ at some $L_\alpha$. This naturally leads to a two
torus of sides $L_5,L_6$. Now, there are
chiral fermions at $(\frac{L_5}{4},\frac{L_6}{4})$,
 $(\frac{3L_5}{4},\frac{L_6}{4})$, $(\frac{L_5}{4},\frac{3L_6}{4})$ and
  $(\frac{3L_5}{4},\frac{3L_6}{4})$.
There are two pairs of Weyl fermions, each pair has two members of opposite
chirality. Together, they make up something like a $u$ and $d$ quark.
The behavior is similar to that of naive fermions where momentum space
has been traded for $x_5$--$x_6$ and we have squared the 2D Dirac operator.

To get open boundaries, like those leading to the simples form
of the overlap, we take, for example,
$v_\alpha^+$ to infinity. The boundary conditions are imposed by this limit
like in the case of the MIT bag model. This procedure certainly works
in the domain wall case. Here, one would guess that an
open square of sides $\frac{L_5}{2}$, $\frac{L_6}{2}$ with four
massless Weyl fermions located at the corners
of the square will be produced. 
We still would have two massless Dirac fermions.

\subsection{MIT-bag boundary conditions}

As just mentioned, additional (but inessential) 
simplifications are possible in the vector-like case.
These simplifications amount to taking the lattice fermion mass parameter
to infinity where-ever this is possible. As a result, fermions are
excluded from some portion of flavor space. This exclusion leaves behind
a boundary and the issue is to determine what boundary condition has
been induced by this process. This problem has been solved many
years ago:

Whenever one uses mass-like terms to exclude a region of space-time
from being visited by fermions one is doing an MIT-bag-like
construction~\cite{mit}. (It is not necessary that the fermion motion
be constrained to a bounded region - the bag can be infinite in 
some directions.) We need to slightly generalize from the original
construction to mass matrices that are
not necessarily diagonal. 

One has a surface of co-dimension 1 and one focuses on
the region close to the surface. The normal to the surface is
$n_\mu$. One wants to exclude fermions from one side of
the surface by taking a mass parameter to infinity there.
On the other side of the surface we want to maintain ordinary
propagation. Call the excluded region ``outside'' and the other
``inside''. In the infinite limit of the mass parameter, a
boundary condition is generated on the inside fermions at the
surface. 

Near the surface we can pick coordinates $\tau$ and $y_i$ 
where $i=1,2,...,d-1$. The $y_i$ are along the surface and the
$\tau$ is perpendicular to it. 
\be
x_\mu = \tau n_\mu + \xi^i_\mu y_i
\ee
where
\be
y\cdot n=0,~~~~ \xi_\mu^i n_\mu=0,~~\xi_\mu^i \xi_\mu^j =\delta^{ij}
\ee
Now,
\be
\partial_\mu = \partial_\tau n_\mu + \xi^i_\mu \partial_{y_i}
\ee
The (Euclidean) equation of motion for the fermions is of the form
\be
(A_\mu \partial_\mu +B)\psi=0
\ee
$A_\mu$ and $B$ are matrices in flavor and spinor space.
$B$ can be written as $B=v\tilde B$ with $\tilde B$ a matrix
with entries of order unity while $v$ is the scale that will go to
infinity. One assumes that $\tilde B$ is invertible. 
Define
\be
\tilde A_\mu = \tilde B^{-1} A_\mu
\ee
For the purpose of determining the boundary conditions
we only care about the derivative terms in the direction
perpendicular to the surface. Hence we need to look only
at
\be
(\tilde A\cdot n \partial_\tau + v )\psi=0
\ee
Writing the solution as
\be
\psi=e^{\frac{1}{z} v\tau} \phi
\ee
we see that $z$ has to be an eigenvalue of $-\tilde A\cdot n$.
By convention, $\tau$ vanishes at the surface and increases 
to positive values outside it. Solutions that tend to diverge
need to be eliminated by the boundary condition. They correspond
to eigenvectors of $-\tilde A\cdot n$ whose real part is positive.
The boundary condition is that $\psi$ be orthogonal to the subspace
spanned by these eigenvectors at the boundary. Note that $\tilde A\cdot n$
will in general not be hermitian or anti-hermitian and we are talking
about its right eigenvectors. 

\section{Generalization of the overlap}

Consider again the Cartesian set-up on the infinite $x_5-x_6$ plane.
Following the overlap construction, our task is to find a formal way
to integrate out all fermions with a fixed four-dimensional gauge background.
This formal way should produce a candidate non-perturbative regularization
of the chiral determinant line bundle we should be familiar with from
continuum field theory. In the overlap case the formal construction admitted
a trivial renormalization and produced a well defined object 
that was subsequently proven to share many topological
and differential features with the 
continuum chiral determinant line bundle. 
The trivial renormalization in the overlap case
amounted to dropping an infinite multiplicative factor from the 
partition function, reducing it to an overlap of two states in a finite
dimensional Hilbert space. We conjecture that here something similar
happens, only that the role of the states is taken by analogues
of Baxter's corner transfer matrices. It is plausible 
that the partition function again contains
infinities only as multiplicative 
constants that are exponentials of local functionals of the 
gauge field background with diverging coefficients.

We use Wilson mass terms to implement the (\ref{mdisc}) structure on
an infinite two dimensional square lattice. This lattice is naturally
divided into four semi-infinite quadrants. Keeping the variables
on the two infinite lines bounding a given quadrant fixed we sum
over all internal fermions. This produces a corner transfer matrix
$K$, for each quadrant, 
depending on the background gauge field and on the quadrant mass
structure. In several cases Baxter has shown that $K$ normalizes
simply to an operator with a discrete spectrum. The spectrum is discrete
since $K$ is, roughly, the exponent of a rotation generator in the plane,
and the rotation is by a finite angle given by $\frac{\pi}{2}$. The generator
of rotations surely has a discrete spectrum. Each $K$ in itself
does not ``know'' that the mass parameters undergo jumps. The
existence of jumps is encoded in the fact that we have four different
$K$-matrices. Extracting the harmless normalization constant which
are infinite, but proportional to the free energy of the very
massive fermions in the gauge field background, we should be left with
an equation of the form:
\be
\label{key}
{\rm chiral~determinant}~=Tr K_1 \{U\} K_2 \{U\}K_3\{U\} K_4\{U\}
\ee
The trace operation performs the integration over all remaining fermion
fields on the boundaries between the quadrants.
This trace is conjectured to produce the bundle we are after 
and requires that an overall phase be left undetermined in the 
process of constructing the $K$-operators. 
The variables $\{U\}$ represent the collection of gauge field
link variables that live on the four dimensional finite toroidal
lattice that replaces the continuum spanned by $x_\mu,~\mu=1,2,3,4$. 

\section{What next ?}
To proceed along this path one would need to do the following:
\begin{itemize}
\item{Show that indeed the matrices $K$ can be constructed in a natural
fashion. The situation is slightly different from that of Baxter in that
the fermions are massive, very far from being 2D conformally invariant. 
However,
the entries of $K$ are, after all, just the inverses of some massive
propagators in a gauge background, so should be essentially local.}
\item{Understand precisely how a natural phase indeterminacy enters
and why~(\ref{key}) should be interpreted as defining a line bundle
over gauge orbit space. Note that if a spectral decomposition for
the $K$ matrices were to apply, one has a nice structure of rings
of states:
\be
\langle\psi_1|\psi_2^\prime\rangle
\langle\psi_2|\psi_3^\prime\rangle
\langle\psi_3|\psi_4^\prime\rangle
\langle\psi_4|\psi_1^\prime\rangle
\ee
}
\item{Find out by what mechanism anomalies are reproduced.}
\item{Take the vector-like case and extract as simple a limit as
possible. In the vector-like case the phase ambiguity should disappear.}
\item{If there is a simple effective action for the 
Dirac case, does
it obey the Ginsparg-Wilson relation~\cite{gw} ?}
\end{itemize}
We end here expressing the hope that some progress 
on the above lines will
occur in the future.

\section{Acknowledgments} I wish to acknowledge useful discussions with
David Kaplan, Seif Randjbar-Daemi and Sasha Zamolodchikov.  
I also wish to acknowledge partial support at the Institute for Advanced Study
in Princeton from a grant in aid by the Monell Foundation, as well as
partial support by the DOE under grant number 
DE-FG02-01ER41165 at Rutgers University.

%


\bigskip

\begin{thebibliography}{99}

\bibitem{plb1} Rajamani Narayanan, Herbert Neuberger, 
Phys. Lett. B302 (1993) 62.

\bibitem{kapplb} David B. Kaplan, Phys. Lett. B288 (1992) 342.

\bibitem{review} H. Neuberger, Ann. Rev. Nucl. Part. Sci 51 (2001) 23.

\bibitem{baxter} R. J. Baxter, J. Stat. Phys 15 (1976) 485;  
R. J. Baxter, J. Stat. Phys 19 (1978) 461; R. J. Baxter,  
J. Stat. Phys 17 (1977) 1.

\bibitem{froslav} S. A.  Frolov, A. A. Slavnov, Phys. Lett. B309 (1993) 344. 



\bibitem{ch} C. G. Callan Jr. and J. A. Harvey, Nucl. Phys. B250 (1985) 427.

\bibitem{ovlap} H. Neuberger, Phys. Lett. B434 (1998) 99; 
H. Neuberger, Phys. Lett. B437 (1998) 117. 


\bibitem{gw} P. Ginsparg, K. Wilson, Phys. Rev. D25 (1982) 2649.

\bibitem{kap} D. B. Kaplan, private communication. 

\bibitem{nagao} K. Nagao, Nucl. Phys. B636 (2002) 264.

\bibitem{mit} A. Chodos, R. L. Jaffe, K. Johnson, C. B. Thorn,
and V. F. Weisskopf, Phys. Rev. D9 (1974) 3471. 


\end{thebibliography}
\end{document}